\documentclass[aps,
12pt,
final,
notitlepage,
oneside,
onecolumn,
nobibnotes,
nofootinbib,
superscriptaddress,
noshowpacs,
centertags]
{revtex4}
\begin{document}
\title{Coherent electroproduction of vector mesons on spinless targets}
\author{\firstname{S.I.}~\surname{Manaenkov}}
\email[]{manaenkov_si@pnpi.nrcki.ru}
\affiliation{B.P.~Konstantinov Petersburg Nuclear Physics Institute of Research Center "Kurchatov Institute",\\
Gatchina 188300, Leningrad region, Russia}
\begin{abstract}
The amplitude ratios of vector-meson production by heavy photons on spinless targets are shown to can be explicitly expressed in terms of the spin-density-matrix elements
(SDMEs) only if the lepton
beam is longitudinally polarized.  Making use of the amplitude ratios as free fit parameters instead of the SDMEs
reduces the number of the real parameters in
data description from 23 to 8. The exact formula for $R = \frac{d \sigma_L}{d t}/\frac{d \sigma_T}{d t}$ in terms of the SDMEs
is obtained for spinless targets and the new approximate formula for $R$ is proposed for nucleon targets.
\end{abstract}

\maketitle

\section{\label{sec:sect-1}Introduction}

Investigation of exclusive vector-meson production in lepton-nucleon deep-inelastic scattering provides information both on production mechanism and on the nucleon 
structure \cite{Sak,Critt,Bauer,INS}. This assumes that incoherent scattering is studied if targets are nuclei 
considered as a system of noninteracting nucleons. Coherent 
scattering from a nucleus when it is intact can give information both
on production mechanism on the nucleus and on its structure. For instance, the leading amplitudes of the vector-meson production on the nucleon at high energies and small 
Bjorken $x_B$ can be used to obtain the gluon distribution in the nucleon  \cite{Rysk93,Rysk99}, while the amplitude  of the coherent vector-meson production on the nucleus can 
provide the information on the gluon distribution in the nucleus. The relation between the gluon distribution in the nucleon and that in the nucleus is
probably not trivial.  Indeed, the unexpected relation between the quark distributions in nuclei and the nucleon was called the EMC effect. It was observed for the first 
time in 1983 \cite{EMC} but up to now, the EMC effect is not ultimately understood (see, for instance, reviews \cite{EMC-rev1, EMC-rev2}).

In the present paper, the simplest case of nuclei with the spin $J=0$ will be considered. All formulas will be obtained in the one-photon-exchange approximation. 
The most popular method of 
phenomenological description of exclusive vector-meson production is the spin-density-matrix-element (SDME) method of the SDME extraction from the fit of the 
angular distribution of final particles to the experimental data. This method was proposed in the pioneering work by Schilling and Wolf \cite{SW} for unpolarized targets and 
generalized to polarized 
targets by Fraas \cite{Fraas} and Diehl \cite{Diehl}. This method which will be called the SDME method considers the SDMEs as free independent parameters. 

The SDMEs were expressed in terms of the amplitudes of the process 
\begin{eqnarray}
\gamma^* (\lambda_{\gamma}) + N(\lambda_{N}) \to V(\lambda_{V})+N (\mu_{N})
\label{gnvn}
\end{eqnarray}
in \cite{SW,Fraas,Diehl}, where $\gamma^*$ denotes the virtual photon created by the beam lepton (or anti-lepton),  $N$ is the nucleon, and $V$ denotes the vector meson. The 
helicities of the particles 
in reaction in Eq.~$(\ref{gnvn})$ are given in parentheses. 
The relation between the SDMEs and the helicity amplitudes is not used in the phenomenological analysis of the data with the SDME method. In the 
amplitude method, the ratios of the helicity amplitudes are extracted directly from the experimental distribution of the final particles. This method was 
for the first time used independently by collaborations H1 \cite{H1-Aar} and HERMES \cite{DC-84}. 

For the spinless nucleus $S$, the helicity amplitudes $F_{\lambda_{V} \lambda_{\gamma}}$ of the reaction
\begin{eqnarray}
\gamma^* (\lambda_{\gamma}) + S \to V(\lambda_{V})+S
\label{gsvs}
\end{eqnarray}
obey the symmetry relations \cite{SW,Diehl}
\begin{eqnarray}
F_{-\lambda_{V} -\lambda_{\gamma}}=(-1)^{\lambda_{V}-\lambda_{\gamma}} F_{\lambda_{V} \lambda_{\gamma}}
\label{par-cons}
\end{eqnarray}
due to parity conservation in strong and electromagnetic interactions. The direct consequence of relations $(\ref{par-cons})$ is the following:
the number of independent helicity amplitudes is 5. We choose the amplitudes $F_{11}=F_{-1-1}$, $F_{10}=-F_{-10}$, $F_{1-1}=F_{-11}$,
$F_{01}=-F_{0-1}$, and $F_{00}$. For high energy and large photon virtuality, the dominant amplitude is $F_{00}$ and convenient choice for the amplitude ratios is 
$F_{\lambda_{V} 
\lambda_{\gamma}}/F_{00}$.
When the photon virtuality, $Q^2$ tends to zero $F_{00} \to 0$, therefore this choice becomes meaningless and the amplitude ratios $F_{\lambda_{V} \lambda_{\gamma}}/F_{11}$
are considered.

\section{Basic equations for the helicity amplitudes}
\label{sect-2}
Let us first define the kinematic variables and reference systems. By definition, the laboratory system (LS) is that where the nuclear target is at rest. The lepton beam energy 
in LS is $E$, while the energy of scattered lepton is $E'$. The four-vector of the initial lepton momentum is $k=(E,{\bf k})$, while the four-momentum of 
the final lepton is $k'=(E',{\bf k}')$, hence the photon four-momentum is $q=k-k'$ and the photon virtuality is $Q^2=-q^2\geq 0$. The photon four-momentum in the 
laboratory 
system looks like $q=(\nu, 0, 0, q_z)$ with the photon 3-momentum aligned along the $Z$-axis in LS. The ratio of 
the virtual photon energy to the 
lepton beam energy in LS will be denoted $y=\nu/E$.
If the target nucleus mass is $M_A$ with $A$ 
being the mass number of the nucleus then the Mandelstam variable for the reaction in Eq.~$(\ref{gsvs})$ is $s=M_A^2-Q^2+2M_A \nu \equiv W^2$, where $W$ denotes the 
center-of-mass (CM) energy of the photon-nucleus system. The variable $x$ for the target nucleus can be defined in LS as $x=Q^2/(2M_A \nu)$, while the Bjorken variable $x_B$
corresponding to scattering from the nucleon with its mass $M_N$ is $x_B=Q^2/(2M_N \nu)$.

The unnormalized spin-density matrix $\Upsilon_{\lambda_V \tilde{\lambda}_V}$ of the produced vector meson is given by the von-Neumann formula
\begin{eqnarray}
\Upsilon_{\lambda_V \tilde{\lambda}_V}=
\sum_{\lambda_{\gamma} \tilde{\lambda}_{\gamma}}  F_{\lambda_{V}  \lambda_{\gamma}}
   F^*_{\tilde{\lambda}_{V} \tilde{\lambda}_{\gamma}} \varrho_{\lambda_{\gamma} \tilde{ \lambda}_{\gamma}},
\label{upsil}
\end{eqnarray}
where $F^*_{\tilde{\lambda}_{V} \tilde{\lambda}_{\gamma}}$ denotes the complex-conjugate amplitude and
$\varrho_{\lambda_{\gamma} \tilde{ \lambda}_{\gamma}}$ is the spin-density matrix of the virtual photon. It contains two terms \cite{SW}
 \begin{eqnarray}
\varrho_{ \lambda_{\gamma}  \tilde{\lambda}_{\gamma}}=
 \varrho^U_{ \lambda_{\gamma}  \tilde{\lambda}_{\gamma}}
  +P_b \varrho^L_{ \lambda_{\gamma}  \tilde{\lambda}_{\gamma}},
\label{gmul}
\end{eqnarray}
where the matrix $\varrho^U_{ \lambda_{\gamma}  \tilde{\lambda}_{\gamma}}$ describes the unpolarized beam case while 
$\varrho^L_{ \lambda_{\gamma} \tilde{\lambda}_{\gamma}}$ corresponds to the additional term due to the longitudinal beam polarization, $P_b$. The matrix 
$\varrho^U_{ \lambda_{\gamma}  \tilde{\lambda}_{\gamma}}$ corresponds to the unit flux of the transversely polarized virtual photons, while 
the ratio of fluxes of the longitudinally to  transversely polarized virtual photons is
\begin{equation}
 \epsilon=\frac{1-y-Q^2/(4E^2)}{1-y+y^2/2+Q^2/(4E^2)}.
\label{def-eps}
\end{equation}
Both matrix $\varrho^U_{ \lambda_{\gamma}  \tilde{\lambda}_{\gamma}}$ and $\varrho^L_{ \lambda_{\gamma}  \tilde{\lambda}_{\gamma}}$can be found in \cite{SW,DC-24}.
They depend on $\epsilon$ and $\Phi$ that denotes the angle between the vector-meson-production plane and 
the lepton-scattering plane (see definition in \cite{DC-24,Joos}). 

To be more specific, we consider $\rho^0$-meson production and the angular distribution of pions from the decay $\rho^0 \to \pi^+ + \pi^-$, though all our results are valid for 
$\phi$-meson production 
followed by the decay $\phi \to K^+K^-$. The angular distribution of the pions in the rest system of 
the $\rho^0$-meson normalized to unity, $\mathcal{W}$ is \cite{DC-24}
\begin{eqnarray} 
\mathcal{N}\mathcal{W}=\sum_{\lambda, \tilde{\lambda}}Y_{1\lambda}(\vartheta, \varphi)Y^*_{1\tilde{\lambda}}(\vartheta, \varphi)\Upsilon_{\lambda 
\tilde{\lambda}},
\label{nw-ups}
\end{eqnarray}
where $\vartheta$ and $\varphi$ are the polar and azimuthal angles of the $\pi^+$-meson 3-momentum in the vector-meson-rest (VMR) system. 
Note, that all the systems under consideration in the present paper are right-handed. The spherical harmonics are chosen in a standard way \cite{Diehl}
\begin{eqnarray}
Y_{1\pm 1}(\vartheta, \varphi) =\mp \sqrt{\frac{3}{8 \pi}}\sin \theta e^{\pm i \varphi},
\label{spharm1}\\
Y_{1 0}(\vartheta, \varphi) =\sqrt{\frac{3}{4 \pi}}\cos \vartheta.
\label{spharm0}
\end{eqnarray}  
Since the normalization condition for the angular distribution looks like
\begin{eqnarray}
\int_{0}^{2 \pi} \frac{d \Phi}{2 \pi} \int_{0}^{\pi}   \sin \vartheta d \vartheta \int_{0}^{2 \pi} d \varphi \mathcal{W}(\Phi, \vartheta, \varphi)=1,
\label{norm-w}
\end{eqnarray}   
the normalization factor $\mathcal{N}$ is defined by 
\begin{eqnarray}
\mathcal{N}=\mathcal{N}_T+ \epsilon \mathcal{N}_L,
\label{norm-nl-nt}\\
\mathcal{N}_T=|F_{11}|^2 +|F_{01}|^2+ |F_{-11}|^2=|F_{11}|^2 +|F_{01}|^2+ |F_{1-1}|^2,
\label{def-nt}\\
\mathcal{N}_L=|F_{10}|^2+ |F_{00}|^2+|F_{-10}|^2
=2|F_{10}|^2+ |F_{00}|^2,
\label{def-nl}
\end{eqnarray}  
where in $(\ref{def-nt})$, $(\ref{def-nl})$ the symmetry property of the helicity amplitudes from $(\ref{par-cons})$ is used.

After substituting into formula $(\ref{nw-ups})$  Eq.~$(\ref{upsil})$, using formula $(\ref{gmul})$ and expressions for the matrix elements of the virtual-photon spin-density 
matrices
$\varrho^U_{ \lambda_{\gamma}  \tilde{\lambda}_{\gamma}}$ and $\varrho^L_{ \lambda_{\gamma}  \tilde{\lambda}_{\gamma}}$ from Refs.~\cite{SW,DC-24} and
spherical harmonics from  $(\ref{spharm1})$, $(\ref{spharm0})$ 
one gets finally the standard formula for the total angular distribution $\mathcal{W}$
where 
\begin{equation}
\mathcal{W}=\mathcal{W}^U+P_b\mathcal{W}^L.
\label{w-total}
\end{equation}
The formulas for $\mathcal{W}^U$ and $\mathcal{W}^L$ with the Schilling-Wolf SDMEs  
$r^{\alpha}_{\lambda_V \tilde{\lambda}_V}$ \cite{SW} presented in \cite{DC-24} are 
the same for scattering both from the nucleons and from the spinless nuclei. But the expressions for the 
Schilling-Wolf SDMEs 
in terms of  the helicity amplitudes are different. In order to obtain the formulas for the spinless targets 
one should first of all put in formulas (77-99) of Ref.~\cite{DC-24}
all the unnatural-parity-exchange (UPE) amplitudes equal to zero. In all the sums with the natural-parity-exchange 
(NPE) amplitudes, the substitution has to be done
\begin{eqnarray}
\widetilde{\sum}\{T_{\lambda_V  \lambda_{\gamma} } T^*_{\lambda'_V \lambda'_{\gamma} }\}
\to F_{\lambda_V \lambda_{\gamma}} F^*_{\lambda'_V \lambda'_{\gamma} }.
\label{t-amp-f}
\end{eqnarray}
The abbreviated notation applied in \cite{DC-24} means
\begin{eqnarray}
\widetilde{\sum}\{T_{\lambda_V  \lambda_{\gamma} } T^*_{\lambda'_V \lambda'_{\gamma} }\}
\equiv \frac{1}{2}\sum_{\lambda_N,\mu_N}\{T_{\lambda_V \mu_N \lambda_{\gamma} \lambda_N} T^*_{\lambda'_V \mu_N \lambda'_{\gamma} \lambda_N}\},
\label{wide-t-amp}\\
\widetilde{\sum}\{U_{\lambda_V  \lambda_{\gamma} } U^*_{\lambda'_V \lambda'_{\gamma} }\}
\equiv \frac{1}{2}\sum_{\lambda_N,\mu_N}\{U_{\lambda_V \mu_N \lambda_{\gamma} \lambda_N} U^*_{\lambda'_V \mu_N \lambda'_{\gamma} \lambda_N}\},
\label{wide-u-amp}
\end{eqnarray}
 where $\lambda_N$ denotes the helicity of the initial nucleon while $\mu_N$ is the final nucleon helicity in the NPE amplitudes 
$T_{\lambda_V \mu_N \lambda_{\gamma} \lambda_N}$ and the UPE amplitudes $U_{\lambda_V \mu_N \lambda_{\gamma} \lambda_N}$.
The normalization factors for the spinless targets in (77-99) of \cite{DC-24} are given by $(\ref{norm-nl-nt}-\ref{def-nl})$.

For the unpolarized SDMEs that determine the angular distribution $\mathcal{W}^U(\Phi, \vartheta,\varphi)$, we have basic equations from which the ratios of the helicity 
amplitudes will be found 
\begin{eqnarray}
\mathcal{N}(1-r^{04}_{00})=  |F_{11}|^2+|F_{1-1}|^2+2 \epsilon |F_{10}|^2,\;\;
\label{d1}\\
\sqrt{2}\mathcal{N}r^{5}_{11}=  \rm{Re}\{(F_{11}-F_{1-1})F_{10}^*\},\;\;
\label{d2}\\
\mathcal{N}r^{1}_{11}=  \rm{Re}\{F_{1-1}F_{11}^*\},\;\;
\label{d3}\\
\mathcal{N}r^{04}_{1-1}=  \rm{Re}\{F_{1-1}F_{11}^*\}-\epsilon |F_{10}|^2,\;\;
\label{d4}\\
\frac{\mathcal{N}}{\sqrt{2}}\{\rm{Im}(r^{6}_{1-1})-r^{5}_{1-1} \}=  \rm{Re}\{F_{10}F_{11}^*\},\;\;   
\label{d5}\\
\frac{\mathcal{N}}{\sqrt{2}}\{\rm{Im}(r^{6}_{1-1})+r^{5}_{1-1} \}=  \rm{Re}\{F_{1-1}F_{10}^*\},\;\;   
\label{d6}\\
\mathcal{N}\{r^{1}_{1-1} -\rm{Im}(r^{2}_{1-1}) \}=  |F_{11}|^2,\;\;
\label{d7}\\
\mathcal{N}\{r^{1}_{1-1} +\rm{Im}(r^{2}_{1-1}) \}=  |F_{1-1}|^2,\;\;
\label{d8}\\
\mathcal{N}r^{04}_{00}=  |F_{01}|^2+\epsilon |F_{00}|^2,\;\;
\label{d9}\\
\frac{\mathcal{N}}{\sqrt{2}}r^{5}_{00}=  \rm{Re}\{F_{00}F_{01}^*\},\;\;
\label{d10}\\
\mathcal{N}r^{1}_{00}=  -|F_{01}|^2,\;\;
\label{d11}\\ 
2\mathcal{N}r^{04}_{10}=  \rm{Re}\{(F_{11}-F_{1-1})F_{01}^*+2 \epsilon F_{10}F_{00}^*\},\;\;
\label{d12}\\
\sqrt{2}\mathcal{N}\{\rm{Re}(r^{5}_{10})-\rm{Im}(r^{6}_{10})\}=  \rm{Re}\{F_{11}F_{00}^*+F_{10}F_{01}^*\},\;\;
\label{d13}\\
\sqrt{2}\mathcal{N}\{\rm{Re}(r^{5}_{10})+\rm{Im}(r^{6}_{10})\}=  \rm{Re}\{F_{10}F_{01}^*-F_{1-1}F_{00}^*\},\;\;
\label{d14}\\
\mathcal{N}\{\rm{Im}(r^{2}_{10})-\rm{Re}(r^{1}_{10})\}=  \rm{Re}\{F_{11}F_{01}^*\},\;\;
\label{d15}\\
\mathcal{N}\{\rm{Im}(r^{2}_{10})+\rm{Re}(r^{1}_{10})\}=  \rm{Re}\{F_{1-1}F_{01}^*\}.\;\;
\label{d16}
\end{eqnarray}

For the polarized SDMEs that determine the angular distribution $\mathcal{W}^L(\Phi, \vartheta,\varphi)$, we have 8 additional basic equations 
\begin{eqnarray}
\sqrt{2}\mathcal{N}r^{8}_{11}=  \rm{Im}\{(F_{11}-F_{1-1})F_{10}^*\},
\label{d17}\\
\mathcal{N}\rm{Im}(r^{3}_{1-1})= -\rm{Im}\{F_{1-1}F_{11}^*\},
\label{d18}\\
\frac{\mathcal{N}}{\sqrt{2}}\{\rm{Im}(r^{7}_{1-1})+r^{8}_{1-1}\}=  -\rm{Im}\{F_{11}F_{10}^*\},
\label{d19}\\
\frac{\mathcal{N}}{\sqrt{2}}\{r^{8}_{1-1}-\rm{Im}(r^{7}_{1-1})\}=  \rm{Im}\{F_{1-1}F_{10}^*\},
\label{d20}\\   
\frac{\mathcal{N}}{\sqrt{2}}r^{8}_{00}=  \rm{Im}\{F_{01}F_{00}^*\},
\label{d21}\\
2\mathcal{N}\rm{Im}(r^{3}_{10})=  \rm{Im}\{(F_{11}+F_{1-1})F_{01}^*\},
\label{d22}\\
\sqrt{2}\mathcal{N}\{\rm{Im}(r^{7}_{10})+\rm{Re}(r^{8}_{10})\}=  \rm{Im}\{F_{11}F_{00}^*-F_{10}F_{01}^*\},
\label{d23}\\
\sqrt{2}\mathcal{N}\{\rm{Im}(r^{7}_{10})-\rm{Re}(r^{8}_{10})\} =  \rm{Im}\{F_{10}F_{01}^*+F_{1-1}F_{00}^*\}.
\label{d24}
\end{eqnarray}  

Equations $(\ref{d1}-\ref{d24})$ for the helicity amplitudes have infinite number of solutions. Indeed, if any set of the amplitudes $F_{\lambda_V \lambda_{\gamma}}$ is a 
solution then 
\begin{eqnarray}
\mathcal{F}_{\lambda_V \lambda_{\gamma}}=\Lambda e^{i \chi}F_{\lambda_V \lambda_{\gamma}} 
\label{scale-phas}
\end{eqnarray}
for arbitrary real parameters $\Lambda \neq 0$ and $\chi$ is also a solution. If we add data on the differential cross section $d \sigma/d t$ of the vector-meson production by 
the virtual photon to the SDME data this fixes $\Lambda$ equal to 1, since \cite{SW}
\begin{eqnarray}
\frac{d \sigma} {d t}=\frac{d \sigma_T} {d t}+\epsilon \frac{d \sigma_L} {d t}=J_{ph}\mathcal{N}=
J_{ph}(\mathcal{N}_T+\epsilon \mathcal{N}_L).
\label{diff-cr-sec}
\end{eqnarray}
Here, $d \sigma_T/d t$ is the differential cross section of the vector-meson production by the transverse photon, while $d \sigma_L/d t$ corresponds to  the longitudinal 
(scalar) 
photon polarization.
The flux factor for the transverse photons in $(\ref{diff-cr-sec})$ is given by \cite{SW}
\begin{eqnarray}
J_{ph}=\frac{1}{32\pi^2(Q^2+\nu^2)}.
\label{j-phot}  
\end{eqnarray}
Measuring $d \sigma/d t$ we experimentally define the value of the normalization factor $\mathcal{N}$ using $(\ref{diff-cr-sec})$ and $(\ref{j-phot})$.
The common phase $\chi$ of all the helicity amplitudes is not fixed by Eqs.~$(\ref{d1}-\ref{d24})$. In the future, we will say that there is a unique solution $F_{\lambda_V 
\lambda_{\gamma}}$ if there is
no solutions $\mathcal{F}_{\lambda_V \lambda_{\gamma}}$ except those obeying Eq.~$(\ref{scale-phas})$ with $\Lambda=1$. Since ratios of the helicity amplitudes
are invariant under transformation in Eq.~$(\ref{scale-phas})$ the meaning of the solution uniqueness for 
the amplitude ratios is usual.  

As is seen from Eqs.~$(38- 39)$ of \cite{DC-24} all SDMEs are multiplied by the functions of the angles $\Phi$, $\vartheta$, $\varphi$
all of which are different from each other. Hence, all $r^{\alpha}_{\lambda_V \tilde{\lambda}_V}$ can be extracted independently.
First of all, we a going to solve the system of Eqs.~$(\ref{d1}-\ref{d16})$ trying to express the amplitudes through $\mathcal{N}$ and $r^{\alpha}_{\lambda_V 
\tilde{\lambda}_V}$.
Note that  $r^{\alpha}_{\lambda_V \tilde{\lambda}_V}$ are dimensionless quantities and may depend on dimensionless amplitude ratios rather than on the helicity
amplitudes themselves. The number of the independent complex amplitude ratios is four that corresponds to 8 real independent functions of $W$, $Q^2$, $t$ for any fixed
vector-meson mass. The total number of the Schilling-Wolf SDMEs is 23. Hence there is 23-8 equations of constraint between the SDMEs. The linear constraint relations follow
immediately from Eqs.~$(77-99)$ of Ref.~\cite{DC-24} adopted to the spinless targets, namely
\begin{eqnarray}
r^{5}_{11}=-r^{5}_{1-1},
\label{u2}\\
r^{8}_{11}=-r^{8}_{1-1},
\label{u3}\\
1-r^{04}_{00}=2(r^{1}_{1-1}+r^{1}_{11}-r^{04}_{1-1}).
\label{u1}
\end{eqnarray}
The nonlinear constraint relations will be presented later. The relations between $r^{\alpha}_{\lambda_V 
\tilde{\lambda}_V}$ are not useless since they permit to
check the quality of the detector property description by Monte-Carlo codes. Indeed, any real detector distorts 
the momentum distribution of the final particles in any process.
Therefore the distributions $\mathcal{W}^U$  and $\mathcal{W}^L$ are obtained when the detector efficiency is taken into account to restore these distributions  from
the experimentally measured distributions of particles in the final state of reaction under study. The detector property is usually described
for modern detectors by  Monte-Carlo codes adopted for the applied detector. If the  used Monte Carlo describes the detector property not perfectly the extracted SDMEs
have systematic errors which lead to the violation of the constraint relations between the  SDMEs $r^{\alpha}_{\lambda_V \tilde{\lambda}_V}$. If the MC codes are good enough
the SDMEs directly extracted from the experimental data with the SDME method should coincide within experimental uncertainties with those calculated with the directly extracted 
helicity-amplitude ratios (obtained by the amplitude method).
A comparison of the SDMEs extracted directly from experimental data and calculated ones with the amplitude ratios
directly extracted from the same data is also the test of a quality of the detector property description with the 
MC codes.

  \section{Vector-meson Production with Unpolarized Beam}
\label{sect-3}
For the unpolarized beam, when $P_b=0$ we have $\mathcal{W}=\mathcal{W}^U$ from Eq.~$(\ref{w-total})$. The main question which must be answered in this paragraph is the 
following:
is it possible to extract the unique set of the amplitude moduli and phase differences from the data on the differential cross section (that fixes $\mathcal{N}$) and the angular 
distribution $\mathcal{W}^U$~? We shall show that the 
system of Eqs.~$(\ref{d1}-\ref{d16})$ has no unique solution, while Eqs.~$(\ref{d1}-\ref{d16})$ and $(\ref{d17}-\ref{d24})$ have the unique solution. 
This means that all moduli and all the phase differences between the amplitudes are single-valued.

It is easy to get formulas for the amplitude moduli. Formulas for $|F_{11}|^2$, $|F_{1-1}|^2$, and $|F_{01}|^2$ follows immediately from $(\ref{d7})$, $(\ref{d8})$, 
and 
$(\ref{d11})$, respectively
\begin{eqnarray}
|F_{11}|^2=\mathcal{N}\{r^{1}_{1-1}-\rm{Im}(r^{2}_{1-1})\},
\label{amp-d7}\\
|F_{1-1}|^2=\mathcal{N}\{r^{1}_{1-1}+\rm{Im}(r^{2}_{1-1})\}.
\label{amp-d8}\\
|F_{01}|^2=-\mathcal{N} r^{1}_{00}.
\label{amp-d11}
\end{eqnarray}
Considering the sum of Eqs.~$(\ref{d9})$ and $(\ref{d11})$ we get
\begin{eqnarray}
|F_{00}|^2=\mathcal{N}\{r^{04}_{00}+r^{1}_{00}\}/\epsilon,
\label{amp-d9d11}
\end{eqnarray}
while the difference of Eqs.~$(\ref{d3})$ and $(\ref{d4})$ gives the last formula of interest
\begin{eqnarray}
|F_{10}|^2=\mathcal{N}\{r^{1}_{11}-r^{04}_{1-1}\}/\epsilon.
\label{amp-d3d4}
\end{eqnarray}
In Eqs.~$(\ref{amp-d7}-\ref{amp-d3d4})$  $\mathcal{N}$ is considered as known from the measurement of the 
differential cross section and using Eqs.~$(\ref{diff-cr-sec})$ and 
$(\ref{j-phot})$. If $d \sigma/d t$ is not measured the 
ratios of the amplitude moduli can be obtained only. For this case, dividing any of Eqs.~$(\ref{amp-d7}-\ref{amp-d3d4})$ by another formula for the amplitude ratio is 
obtained. For instance, 
dividing Eq.~$(\ref{amp-d7})$ by Eq.~$(\ref{amp-d9d11})$ one gets 
\begin{eqnarray}
\frac{|F_{11}|^2}{|F_{00}|^2}=\frac{\epsilon\{r^{1}_{1-1}-\rm{Im}(r^{2}_{1-1})\}}{r^{04}_{00}+r^{1}_{00}}
\label{(f11/f00)^2}
\end{eqnarray}
and so on.
 
The phase differences between the amplitudes cannot be obtained single-valued since Eqs.~$(\ref{d1}-\ref{d16})$ contain only real parts of the amplitude products of the 
type $F_{\lambda_V \lambda_{\gamma}}F^*_{\tilde{\lambda}_V \tilde{\lambda}_{\gamma}}$ which contain only the cosines of the phase differences.  Indeed, the  real part of the 
product of any two complex numbers $F_{\lambda_V \lambda_{\gamma}}$ and 
$F^*_{\tilde{\lambda}_V \tilde{\lambda}_{\gamma}}$ is
\begin{equation}
\rm{Re} \{F_{\lambda_V \lambda_{\gamma}}F^*_{\tilde{\lambda}_V \tilde{\lambda}_{\gamma}}\}=|F_{\lambda_V \lambda_{\gamma}}| |F_{\tilde{\lambda}_V \tilde{\lambda}_{\gamma}}| 
\cos (\delta_{\lambda_V \lambda_{\gamma}}-\delta_{\tilde{\lambda}_V \tilde{\lambda}_{\gamma}}),
\label{real-z1-z2}
\end{equation}
where $\delta_{\tilde{\lambda}_V \tilde{\lambda}_{\gamma}}$ denotes the phase of the complex number $F_{\lambda_V \lambda_{\gamma}}$, while  $|F_{\lambda_V \lambda_{\gamma}}|$ 
is 
the modulus of $F_{\lambda_V \lambda_{\gamma}}$. If a set of the phases $\delta_{\lambda_V \lambda_{\gamma}}$ 
of all the amplitudes $F_{\lambda_V \lambda_{\gamma}}$ is a solution 
of Eqs.~$(\ref{d1}-\ref{d16})$,  then the set with $-\delta_{\lambda_V \lambda_{\gamma}}$ is also the solution of the same equations. Hence at least two solutions of basic 
equations $(\ref{d1}-\ref{d16})$ exist. Considering Eqs.~$(\ref{d17}-\ref{d24})$ which contain imaginary parts of the bilinear amplitude products $F_{\lambda_V 
\lambda_{\gamma}}F^*_{\tilde{\lambda}_V \tilde{\lambda}_{\gamma}}$ 
we can get sines of the phase differences since
\begin{equation}  
\rm{Im} \{F_{\lambda_V \lambda_{\gamma}}F^*_{\tilde{\lambda}_V \tilde{\lambda}_{\gamma}}\}=|F_{\lambda_V \lambda_{\gamma}}| |F_{\tilde{\lambda}_V \tilde{\lambda}_{\gamma}}|
\sin (\delta_{\lambda_V \lambda_{\gamma}}-\delta_{\tilde{\lambda}_V \tilde{\lambda}_{\gamma}}).
\label{imag-z1-z2}
\end{equation}
Adding Eqs.~$(\ref{d17}-\ref{d24})$ to Eqs.~$(\ref{d1}-\ref{d16})$ makes the phase differences single-valued. 

Let us try to find any solution of Eqs.~$(\ref{d1}-\ref{d16})$ for cosines of phase differences using equation $(\ref{real-z1-z2})$. Substituting Eqs.~$(\ref{d15})$
for $\rm{Re} \{F_{11}F^*_{01}\}$ and $(\ref{amp-d7})$, $(\ref{amp-d11})$ respectively for moduli of $F_{11}$ and $F_{01}$ into $(\ref{real-z1-z2})$ one gets
\begin{equation}
\cos (\delta_{11}-\delta_{01})=\frac{\rm{Im}(r^{2}_{10})-\rm{Re}(r^{1}_{10})}{\sqrt{\{\rm{Im}(r^{2}_{1-1}) -r^{1}_{1-1}\} r^{1}_{00}}}.\;\;
\label{cos-del-11-01}
\end{equation}
In an analogous way, substituting Eqs.~$(\ref{d16})$
for $\rm{Re} \{F_{1-1}F^*_{01}\}$ and $(\ref{amp-d8})$, $(\ref{amp-d11})$ respectively for moduli of $F_{1-1}$ and $F_{01}$ into $(\ref{real-z1-z2})$ one gets
\begin{equation}
\cos (\delta_{1-1}-\delta_{01})=\frac{\rm{Im}(r^{2}_{10})+\rm{Re}(r^{1}_{10})}{\sqrt{-\{\rm{Im}(r^{2}_{1-1}) +r^{1}_{1-1}\} r^{1}_{00}}},\;\;
\label{cos-del-1m1-01}
\end{equation}
while substituting Eqs.~$(\ref{d3})$
for $\rm{Re} \{F_{1-1}F^*_{11}\}$ and $(\ref{amp-d7})$, $(\ref{amp-d8})$ respectively for moduli of $F_{11}$ and $F_{1-1}$ into $(\ref{real-z1-z2})$ one 
obtains
\begin{equation}
\cos (\delta_{11}-\delta_{1-1})=\frac{r^{1}_{11}}{\sqrt{\{r^{1}_{1-1}\}^2-\{\rm{Im}(r^{2}_{1-1})\}^2}}.\;\;
\label{cos-del-11-1m1}
\end{equation}
Since $(\delta_{11}-\delta_{1-1})=(\delta_{11}-\delta_{01})-(\delta_{1-1}-\delta_{01})$ than according to the basic trigonometric formula, we have the equation
\begin{equation}
\cos (\delta_{11}-\delta_{1-1})=\cos (\delta_{11}-\delta_{01})\cos (\delta_{1-1}-\delta_{01})+\sin (\delta_{11}-\delta_{01})\sin (\delta_{1-1}-\delta_{01}).
\label{cos-del-11-1m1=sum}
\end{equation}
It is obvious from $(\ref{cos-del-11-1m1=sum})$ that
\begin{eqnarray}
\nonumber
\{\cos (\delta_{11}-\delta_{1-1})-\cos (\delta_{11}-\delta_{01})\cos (\delta_{1-1}-\delta_{01})\}^2=\\=\sin ^2 (\delta_{11}-\delta_{01})\sin ^2(\delta_{1-1}-\delta_{01})
=\{1-\cos ^2 (\delta_{11}-\delta_{01})\}\{1-\cos ^2(\delta_{1-1}-\delta_{01})\}.
\label{dif1-fin}
\end{eqnarray}
Expressing all cosines in $(\ref{dif1-fin})$ with the help of $(\ref{cos-del-11-01})$, $(\ref{cos-del-1m1-01})$, and $(\ref{cos-del-11-1m1})$ we obtain the first non-linear
constraint equation
\begin{eqnarray}
\nonumber
\Bigl \{r^{1}_{11} |r^{1}_{00}|+\Bigl( \rm{Re}(r^{1}_{10})\Bigr)^2-\Bigl(\rm{Im}(r^{2}_{10})\Bigr)^2 \Bigr \}^2\\
\nonumber
=\Bigl\{\Bigl(r^{1}_{1-1}-\rm{Im}(r^{2}_{1-1})\Bigr)r^{1}_{00}+\Bigl(\rm{Im}(r^{2}_{10})-\rm{Re}(r^{1}_{10})\Bigr)^2 \Bigr \} \\ 
 \times \Bigl\{\Bigl(r^{1}_{1-1}+\rm{Im}(r^{2}_{1-1})\Bigr)r^{1}_{00}+\Bigl(\rm{Im}(r^{2}_{10})+\rm{Re}(r^{1}_{10})\Bigr)^2  \Bigr\}.
\label{first-non-lin}
\end{eqnarray}

Let us find $\cos(\delta_{11}-\delta_{10})$ and $\cos(\delta_{1-1}-\delta_{10})$ respectively from Eqs.~$(\ref{d5})$, $(\ref{amp-d7})$, $(\ref{amp-d3d4})$  and 
Eqs.~$(\ref{d6})$, $(\ref{amp-d8})$, $(\ref{amp-d3d4})$, 
while $\cos(\delta_{11}-\delta_{1-1})$ from $(\ref{cos-del-11-1m1})$.  
Since  $(\delta_{11}-\delta_{1-1})=(\delta_{11}-\delta_{10})-(\delta_{1-1}-\delta_{10})$ we can obtain the second non-linear constraint relation 
\begin{eqnarray}
\nonumber
\Bigl \{r^{1}_{11} \Bigl (r^{1}_{11}-r^{04}_{1-1}\Bigr )-\frac{\epsilon}{2}\Bigl [\Bigl(\rm{Im}(r^{6}_{1-1})\Bigr)^2 -
\Bigl( r^{5}_{1-1}\Bigr)^2\Bigr]  \Bigr \}^2\\
\nonumber
=\Bigl\{\Bigl(r^{1}_{1-1}-\rm{Im}(r^{2}_{1-1})\Bigr) \Bigl(r^{1}_{11}- r^{04}_{1-1}\Bigr ) -\frac{\epsilon}{2} \Bigl(\rm{Im}(r^{6}_{1-1})-r^{5}_{1-1}\Bigr)^2 \Bigr \} 
\\
 \times \Bigl\{\Bigl(r^{1}_{1-1}+\rm{Im}(r^{2}_{1-1})\Bigr)\Bigl(r^{1}_{11}- r^{04}_{1-1}\Bigr ) -\frac{\epsilon}{2}\Bigl(\rm{Im}(r^{6}_{1-1})+r^{5}_{1-1}\Bigr)^2 \Bigr \}
\label{second-non-lin}
\end{eqnarray}
in an analogous way as
Eq.~$(\ref{first-non-lin})$ was obtained and so on.
\section{Vector-meson Production with Polarized Beam}
\label{sect-4}
For polarized beam, the angular distribution  $\mathcal{W}^L$ is added to $\mathcal{W}^U$ in Eq.~$(\ref{w-total})$.
From Eqs.~$(\ref{d1}-\ref{d16})$ and $(\ref{d17}-\ref{d24})$, the relations for complex numbers follow, namely, relation
\begin{eqnarray}
F_{1-1}F_{11}^*=\mathcal{N}\Bigl\{r^{1}_{11}-i\rm{Im}(r^{3}_{1-1})\Bigr\}
\label{f1m1f11}
\end{eqnarray}
can be obtained from  Eqs.~$(\ref{d3})$ and $(\ref{d18})$, while formula
\begin{eqnarray}
F_{01}F_{00}^*=\frac{\mathcal{N}}{\sqrt{2}}\Bigl\{r^{5}_{00}+ir^{8}_{00}\Bigr\}
\label{f01f00}
\end{eqnarray}
follows from Eqs.~$(\ref{d10})$ and $(\ref{d21})$.
If we combine Eq.~$(\ref{d2})$ with Eq.~$(\ref{d5})$ to get a real part of amplitude products and Eq.~$(\ref{d17})$ with 
$(\ref{d20})$ to get the imaginary part of the amplitude products we obtain easily
\begin{eqnarray}
(F_{11}+F_{1-1})F_{10}^* 
=\mathcal{N}\sqrt{2}\Bigl\{\rm{Im}(r^{6}_{1-1})-r^{5}_{1-1}-r^{5}_{11} +i\Bigl(r^{8}_{11}+r^{8}_{1-1}-\rm{Im}(r^{7}_{1-1}) \Bigr)\Bigr\}.
\label{f11+f1m1f10--1}
\end{eqnarray}
If we take into account in Eq.~$(\ref{f11+f1m1f10--1})$ relations $(\ref{u2})$ and $(\ref{u3})$ we simplify it and obtain
\begin{eqnarray}
(F_{11}+F_{1-1})F_{10}^*
=\mathcal{N}\sqrt{2}\Bigl\{\rm{Im}(r^{6}_{1-1}) -i\rm{Im}(r^{7}_{1-1})\Bigr\},
\label{f11+f1m1f10}
\end{eqnarray}
while relation 
\begin{eqnarray}
(F_{11}+F_{1-1})F_{01}^*=2\mathcal{N}\Bigl\{\rm{Im}(r^{2}_{10})+i \rm{Im}(r^{3}_{10})\Bigr\}
\label{f11+f1m1f01}
\end{eqnarray}
follows from the sum of Eqs.~$(\ref{d15})$, $(\ref{d16})$ and also $(\ref{d22})$. 
Calculating the difference of Eqs.~$(\ref{d13})$,  $(\ref{d14})$ and the sum of Eqs.~$(\ref{d23})$, $(\ref{d24})$ 
the formula
\begin{eqnarray}
(F_{11}+F_{1-1})F_{00}^*=\mathcal{N}\sqrt{8}\Bigl\{-\rm{Im}(r^{6}_{10})+i\rm{Im}(r^{7}_{10})\Bigr\}
\label{f11+f1m1f00}
\end{eqnarray}
can be obtained. Considering the difference of Eqs.~$(\ref{d5})$ and $(\ref{d2})$ one obtains the real part while Eq.~$(\ref{d20})$ gives the imaginary part of the product
$F_{1-1}F_{10}^*$, therefore
\begin{eqnarray}
F_{1-1}F_{10}^*=\frac{\mathcal{N}}{\sqrt{2}}\Bigl\{\rm{Im}(r^{6}_{1-1})-r^{5}_{1-1}-2r^{5}_{11}
+i\Bigl(r^{8}_{1-1} -\rm{Im}(r^{7}_{1-1})\Bigr)\Bigr\}.
\label{f1m1f10-01}
\end{eqnarray}
The equation
\begin{eqnarray}
F_{11}F_{10}^*=\frac{\mathcal{N}}{\sqrt{2}}\Bigl\{\rm{Im}(r^{6}_{1-1})-r^{5}_{1-1}
+i\Bigl(2r^{8}_{11}+ r^{8}_{1-1}-\rm{Im}(r^{7}_{1-1})\Bigr)\Bigr\}
\label{f11f10-01}  
\end{eqnarray}
follows immediately from  Eqs.~$(\ref{d5})$ and the sum of $(\ref{d17})$, $(\ref{d20})$. Substitution of Eq.~$(\ref{u2})$ into $(\ref{f1m1f10-01})$
and Eq.~$(\ref{u3})$ into $(\ref{f11f10-01})$ leads respectively to the more simple formulas
\begin{eqnarray}
F_{1-1}F_{10}^*=\frac{\mathcal{N}}{\sqrt{2}}\Bigl\{\rm{Im}(r^{6}_{1-1})-r^{5}_{11}
+i\Bigl(r^{8}_{1-1} -\rm{Im}(r^{7}_{1-1})\Bigr)\Bigr\},
\label{f1m1f10}\\
F_{11}F_{10}^*=\frac{\mathcal{N}}{\sqrt{2}}\Bigl\{\rm{Im}(r^{6}_{1-1})-r^{5}_{1-1}
+i\Bigl(r^{8}_{11}-\rm{Im}(r^{7}_{1-1})\Bigr)\Bigr\}.
\label{f11f10}
\end{eqnarray}  
 
Dividing  Eq.~$(\ref{f01f00})$ by Eq.~$(\ref{amp-d9d11})$ we get the first amplitude ratio
\begin{eqnarray}
\frac{F_{01}}{F_{00}} \equiv \frac{F_{01}F_{00}^*}{F_{00}F_{00}^*}
\equiv
\frac{F_{01}F_{00}^*}{|F_{00}|^2}=
\frac{\epsilon\bigl\{r^{5}_{00}+ir^{8}_{00}\bigr\}}{\sqrt{2}\bigl\{r^{04}_{00}+r^{1}_{00}\bigr\}}.
\label{f01-f00}
\end{eqnarray}
Dividing Eq.~$(\ref{f11+f1m1f10})$ by Eq.~$(\ref{f11+f1m1f00})$ and considering the complex conjugate of the SDME-linear-combination ratio we obtain the second  amplitude ratio
\begin{eqnarray}
 \frac{F_{10}}{F_{00}}=-\frac{\rm{Im}(r^{6}_{1-1})+i\rm{Im}(r^{7}_{1-1})}{2\bigl\{\rm{Im}(r^{6}_{10})+i\rm{Im}(r^{7}_{10})\bigr\}}.
\label{f10-f00}
\end{eqnarray}
Dividing Eq.~(\ref{f1m1f11}) by (\ref{amp-d7}) we get the ratio $F_{1-1}/F_{11}$
\begin{eqnarray}
\frac{F_{1-1}}{F_{11}}\equiv \frac{F_{1-1}F_{11}^*}{|F_{11}|^2}=\frac{r^{1}_{11}-i\rm{Im}(r^{3}_{1-1})}{\bigl\{r^{1}_{1-1}-\rm{Im}(r^{2}_{1-1})\bigr\}}.
\label{f1m1-f11}
\end{eqnarray}
Dividing Eq.~(\ref{f11+f1m1f00}) by (\ref{amp-d9d11}) one gets a chain of equalities
\begin{eqnarray}
\nonumber
 \frac{(F_{11}+F_{1-1})F_{00}^*}{|F_{00}|^2} 
\equiv\Bigl \{1+\frac{F_{1-1}}{F_{11}}\Bigr \}  \frac{F_{11}F_{00}^*}{|F_{00}|^2}\equiv\\
\equiv\Bigl \{1+\frac{F_{1-1}}{F_{11}}\Bigr \}  \frac{F_{11}}{F_{00}} =
\frac{\epsilon\sqrt{8} \bigl\{-\rm{Im}(r^{6}_{10})+i \rm{Im}(r^{7}_{10})\bigr \}}{r^{04}_{00}+r^{1}_{00}}.
\label{ex-f11-f00}
\end{eqnarray}  
Substituting the ratio $F_{1-1}/F_{11}$ given by auxiliary Eq.~(\ref{f1m1-f11}) in Eq.~(\ref{ex-f11-f00}) we rewrite  Eq.~(\ref{ex-f11-f00}) as
\begin{eqnarray}
\Bigl \{1+\frac{r^{1}_{11}-i\rm{Im}(r^{3}_{1-1})}{r^{1}_{1-1}-\rm{Im}(r^{2}_{1-1})} \Bigr\}  \frac{F_{11}}{F_{00}} 
=\frac{\epsilon\sqrt{8} \bigl\{-\rm{Im}(r^{6}_{10})+i \rm{Im}(r^{7}_{10})\bigr \}}{r^{04}_{00}+r^{1}_{00}}.
\label{ex1-f11-f00}
\end{eqnarray}  
We obtain finally from Eq.~$(\ref{ex1-f11-f00})$ that
\begin{eqnarray}
 \frac{F_{11}}{F_{00}}=
\frac{\epsilon\sqrt{8} \bigl\{-\rm{Im}(r^{6}_{10})+i \rm{Im}(r^{7}_{10})\bigr \}\bigl\{ r^{1}_{1-1}-\rm{Im}(r^{2}_{1-1})\bigr \}}
{\bigl\{r^{04}_{00}+r^{1}_{00}\bigr \}\bigl\{r^{1}_{11}+r^{1}_{1-1}-\rm{Im}(r^{2}_{1-1}) -i \rm{Im}(r^{3}_{1-1})\bigr \}}.
\label{f11-f00}
\end{eqnarray}  
Multiplying (\ref{f1m1-f11}) and (\ref{f11-f00}) we get the last linearly independent ratio of the helicity amplitudes to $F_{00}$
\begin{eqnarray}
 \frac{F_{1-1}}{F_{00}}=\frac{\epsilon\sqrt{8} \bigl\{-\rm{Im}(r^{6}_{10})+i \rm{Im}(r^{7}_{10})\bigr \}\bigl\{ r^{1}_{11}-i\rm{Im}(r^{3}_{1-1}) \bigr \}}
{\bigl\{r^{04}_{00}+r^{1}_{00}\bigr \}\bigl\{r^{1}_{11}+r^{1}_{1-1}-\rm{Im}(r^{2}_{1-1}) -i \rm{Im}(r^{3}_{1-1})\bigr \}}.
\label{f1m1-f00}
\end{eqnarray}

Formulas $(\ref{f01-f00})$, $(\ref{f10-f00})$, $(\ref{f11-f00})$, $(\ref{f1m1-f00})$ provide ratios of the amplitudes to $F_{00}$, 
which are convenient for deep-inelastic scattering since $F_{00}$ dominates at $Q^2 \to \infty$ and is non-zero at $v_T=0$. When $Q^2 \to 0$ these ratios become meaningless
since $F_{00} \to 0$. The $S$-Channel Helicity Conservation (SCHC) approximation assumes that nonzero amplitudes  
at $v_T=0$ are only amplitudes with $\lambda_V=\lambda_{\gamma}$.
It is convenient to divide all helicity amplitudes by another SCHC amplitude which is nonzero at $Q^2=0$. 
It is the amplitude $F_{11}$. The first such ratio is given by Eq.~$(\ref{f1m1-f11})$, while
the ratio $F_{00}/F_{11}$ gives the formula inverse to Eq.~$(\ref{f11-f00})$. Dividing the relation complex-conjugate 
to Eq.~$(\ref{f11f10})$ by Eq.~$(\ref{amp-d7})$, we get the final result for $F_{10}/F_{11}$
\begin{eqnarray}
 \frac{F_{10}}{F_{11}}=\frac{F_{10}F_{11}^*}{|F_{11}|^2} 
=\frac{\rm{Im}(r^{6}_{1-1})-r^{5}_{1-1}-i\bigl(r^{8}_{11}-\rm{Im}(r^{7}_{1-1}) \bigr)}{\sqrt{2}\bigl\{r^{1}_{1-1}-\rm{Im}(r^{2}_{1-1})\bigr\}}.
\label{f10-f11} 
\end{eqnarray}
Analogously, dividing the relation complex-conjugate to Eq.~$(\ref{f11+f1m1f01})$ by Eq.~$(\ref{amp-d7})$, we get 
\begin{eqnarray}
 \frac{F_{01}(F_{11}+F_{1-1})^*}{|F_{11}|^2} 
=\frac{F_{01}}{F_{11}}\Bigl [1+\Bigl(\frac{F_{1-1}}{F_{11}}\Bigr)^* \Bigr ]= \frac{2\bigl\{\rm{Im}(r^{2}_{10})-i \rm{Im}(r^{3}_{10})\bigr\}}{r^{1}_{1-1}-\rm{Im}(r^{2}_{1-1})}.
\label{ex-f01-f11}
\end{eqnarray}
Substituting   for $F_{1-1}/F_{11}$  in the square brackets in Eq.~$(\ref{ex-f01-f11})$ formula 
$(\ref{f1m1-f11})$ we obtain for $F_{01}/F_{11}$ the final result
\begin{eqnarray}
 \frac{F_{01}}{F_{11}}=
\frac{2\bigl\{\rm{Im}(r^{2}_{10})-i \rm{Im}(r^{3}_{10})\bigr\}}{r^{1}_{11}+r^{1}_{1-1}-\rm{Im}(r^{2}_{1-1})+i\rm{Im}(r^{3}_{1-1})}.
\label{f01-f11}
\end{eqnarray}
\section{Virtual-photon Longitudinal-to-transverse Cross-section Ratio}
\label{sect-5}
The ratio of the differential cross sections of vector-meson production by longitudinally-polarized ($\frac{d \sigma _L}{d t}$) to transversely-polarized 
($\frac{d \sigma _T}{d t}$) virtual photons is equal to \cite{SW}  
\begin{eqnarray}
R \equiv \frac{d \sigma _L}{ d t}\Big / \frac{d \sigma _T}{ d t}=\frac{\mathcal{N}_L}{\mathcal{N}_T} 
=\frac{2|F_{10}|^2+ |F_{00}|^2}{|F_{11}|^2+ |F_{01}|^2 +|F_{1-1}|^2}.
\label{def-sl-st}
\end{eqnarray}  
Here, Eqs.~$(\ref{def-nt})$, $(\ref{def-nl})$, and $(\ref{diff-cr-sec})$ are used. Substituting formulas for the moduli of the helicity amplitudes 
$(\ref{amp-d7}-\ref{amp-d3d4})$ 
into 
Eq.~$(\ref{def-sl-st})$ we get
\begin{eqnarray}
R=\frac{r^{04}_{00}+r^{1}_{00}+2(r^{1}_{11}-r^{04}_{1-1})}{\epsilon(2r^{1}_{1-1}-r^{1}_{00})}.
\label{r-less}
\end{eqnarray}
We conclude from Eq.~$(\ref{r-less})$ that the formula for the 
ratio of the cross sections, $R$ in terms of the SDMEs for  spinless targets  is exact, therefore we do not need the Rosenbluth decomposition \cite{Ros} to extract $R$ from the 
experimental data. In order to obtain the cross sections  $\frac{d \sigma _T}{d t}$ and $\frac{d \sigma _L}{d t}$ from 
the measured value of $\frac{d \sigma }{d t}=\frac{d \sigma _T}{d t}+\epsilon \frac{d \sigma _L}{d t}$ data for the same $\nu$, $Q^2$, $t$  
but for at least two beam energies (and hence two values of $\epsilon$ according to Eq.~$(\ref{def-eps})$) are needed. 
We remind  that the only way to extract the exact value for $R$ from the electroproduction data on the nucleon is to use the Rosenbluth decomposition, while any formula for $R$ 
in terms of the 
SDMEs, obtained at one beam energy, is approximate. 

Using constraint relation $(\ref{u1})$ it is possible to rewrite Eq.~$(\ref{r-less})$ in another form, namely
\begin{eqnarray}
R=\frac{1}{\epsilon}\Bigl\{\frac{1}{(2r^{1}_{1-1}-r^{1}_{00})}-1\Bigr\}.
\label{r-less-upe}
\end{eqnarray}
Formulas $(\ref{r-less})$ and $(\ref{r-less-upe})$ are to give the same values for $R$ withing experimental uncertainties. Since there are only two SDMEs in the expression
for $R$ in Eq.~$(\ref{r-less-upe})$ the cross-section ratio has usually the better accuracy than $R$ defined by Eq.~$(\ref{r-less})$ which contains five SDMEs. We conclude 
that for mutually dependent SDMEs there is no unique formula for $R$. Various formulas for $R$ may express it in terms of SDME sets different from each other and these $R$ can 
have different experimental uncertainties.

Originally, the SDMEs in the Schilling-Wolf representation were used for the vector-meson production on the nucleons.
The simplest idea is to use Eq.~$(\ref{r-less})$ or Eq.~$(\ref{r-less-upe})$ for an improvement of the description of the virtual-photon 
longitudinal-to-transverse cross-section ratio on the nucleon.
In order to distinguish observables for the nucleon from the same quantities for spinless targets we will mark 
the former by tilde.
Using formulas for the SDMEs $\widetilde{r}_{\lambda_V \tilde{\lambda}_V}^{\alpha}$  (see \cite{SW,DC-24}) 
in terms of the helicity amplitudes $T_{\lambda_V \mu_N \lambda_{\gamma} \lambda_N}$ 
and $U_{\lambda_V \mu_N \lambda_{\gamma} \lambda_N}$ we obtain the formula for the virtual-photon 
longitudinal-to-transverse cross-section ratio on the nucleon from Eq.~$(\ref{r-less})$ 
\begin{eqnarray}
\nonumber
\widetilde{R}_U=
\frac{\widetilde{r}^{04}_{00}+\widetilde{r}^{1}_{00}+2(\widetilde{r}^{1}_{11}-\widetilde{r}^{04}_{1-1})}
{\epsilon(2\widetilde{r}^{1}_{1-1}-\widetilde{r}^{1}_{00})}=\\=
\frac{\widetilde{\sum}\bigl\{\epsilon \bigl (|T_{00}|^2+2|T_{10}|^2 
-2|U_{10}|^2 \bigr )
+2|U_{01}|^2+4\rm{Re}\bigl (U_{1-1}U_{11}^*\bigr )\bigr \}}
{\epsilon \widetilde{\sum}\bigl \{|T_{11}|^2+|T_{01}|^2+|T_{-11}|^2 
-|U_{11}|^2-|U_{01}|^2-|U_{-11}|^2 \bigr \}}.
\label{rs-amp}
\end{eqnarray}
The formulas for the true ratio of the cross sections on the nucleon are \cite{SW,DC-24}
\begin{eqnarray}
\widetilde{R}=\widetilde{\mathcal{N}}_L/\widetilde{\mathcal{N}}_T, 
\label{r-nl-nt-nucl}\\
\widetilde{\mathcal{N}}_L=\widetilde{\sum} \bigl (|T_{00}|^2+2|T_{10}|^2+ 2|U_{10}|^2\bigr ),\;\;
\label{nl-1/2}\\
\widetilde{\mathcal{N}}_T=\widetilde{\sum} \bigl (|T_{11}|^2+|T_{01}|^2+ |T_{-11}|^2 
+ |U_{11}|^2+|U_{01}|^2+ |U_{-11}|^2 \bigr ),\;\;\;
\label{nt-1/2}\\
\widetilde{\mathcal{N}}=\widetilde{\mathcal{N}}_T+\epsilon \widetilde{\mathcal{N}}_L.\;\;\;
\label{n=nt-exnl}
\end{eqnarray}
A comparison of $\widetilde{R}_U$ given by Eq.~$(\ref{rs-amp})$ and the true ratio $\widetilde{R}$ defined by Eqs.~$(\ref{r-nl-nt-nucl}-\ref{nt-1/2})$ shows that the 
contributions
of the UPE-amplitudes are taken into account in Eq.~$(\ref{rs-amp})$ incorrect. For the collider experiments at very high energies, the contribution of 
the UPE-amplitudes is strongly suppressed (as usually expected in Regge phenomenology \cite{KS,IW})
then $\widetilde{R}_U=\widetilde{R}$ with a good accuracy. But for fixed-target experiments, a neglect of the UPE-amplitudes
especially the greatest among them, $U_{1\frac{1}{2}1\frac{1}{2}}$ can give an appreciable discrepancy. For instance, the contribution of $U_{1\frac{1}{2}1\frac{1}{2}}$ is a 
measurable effect at $W \approx 6$ GeV for $\rho^0$-meson production at HERMES \cite{DC-24,DC-84} and is crucially important for $\omega$-production \cite{DC-95,DC-95-err}.

Now, we are going to improve Eq.~$(\ref{rs-amp})$ for $\widetilde{R}_U$. Let us consider the numerator of Eq.~$(\ref{rs-amp})$ and express it in terms of the 
helicity amplitudes
using standard equations for $\widetilde{r}_{\lambda_V \tilde{\lambda}_V }^{\alpha}$ from \cite{SW,DC-24}
\begin{eqnarray}
\nonumber
D_S \equiv \widetilde{r}^{04}_{00}+\widetilde{r}^{1}_{00}+2(\widetilde{r}^{1}_{11}-\widetilde{r}^{04}_{1-1})=\\
\nonumber
=\{\epsilon \widetilde{\mathcal{N}}_L+  
\widetilde{\sum}[-4\epsilon|U_{10}|^2
+2|U_{01}|^2+4\rm{Re}\bigl (U_{1-1}U_{11}^* \bigr)]\}/\widetilde{\mathcal{N}}=\\
=\frac{\epsilon \widetilde{\mathcal{N}}_L}{\widetilde{\mathcal{N}}}+\beta_D=\frac{1}{1+1/(\epsilon \widetilde{R})}+\beta_D,   
\label{rs-numera}
\end{eqnarray}
where $\widetilde{\mathcal{N}}_L$,  $\widetilde{\mathcal{N}}_T$, and 
$\widetilde{\mathcal{N}}$ are defined by
Eqs.~$(\ref{nl-1/2}-\ref{n=nt-exnl})$, while $\widetilde{R}$ is given by Eq.~$(\ref{r-nl-nt-nucl})$. The quantity $\beta_D$ according to Eq.~$(\ref{rs-numera})$  is  
\begin{eqnarray}
\beta_D=\bigl\{\widetilde{\sum}[-4\epsilon|U_{10}|^2+2|U_{01}|^2 
+4\rm{Re}\bigl (U_{1-1}U_{11}^* \bigr )]\bigr\}/\widetilde{\mathcal{N}}.
\label{def-beta-u}
\end{eqnarray}
The exact formula for $\widetilde{R}$ follows from the right-hand side of Eq.~$(\ref{rs-numera})$
\begin{eqnarray}
\widetilde{R}=\frac{1}{\epsilon}\frac{D_S-\beta_D}{1-D_S+\beta_D}.
\label{ds-r-beta-prec}
\end{eqnarray}
The parameter $\beta_D$ at high energy is small since it is the ratio of bilinear products of the UPE-amplitudes to $\widetilde{\mathcal{N}}$ which contains the sum of squared 
moduli of all NPE- and UPE-amplitudes. We would like to stress that the numerator in formula $(\ref{def-beta-u})$ for $\beta_D$ does not contain the modulus square of the 
largest UPE-amplitude $U_{1\frac{1}{2}1\frac{1}{2}}$.
Neglecting the value of $\beta_D$ in Eq.~$(\ref{ds-r-beta-prec})$ one obtains the approximate formula of interest
\begin{eqnarray}
\widetilde{R} \approx \widetilde{R}_D=\frac{1}{\epsilon}\frac{D_S}{1-D_S} 
=\frac{\widetilde{r}^{04}_{00}+\widetilde{r}^{1}_{00}+2(\widetilde{r}^{1}_{11}-\widetilde{r}^{04}_{1-1})}{\epsilon 
\{1-\widetilde{r}^{04}_{00}-\widetilde{r}^{1}_{00}-2(\widetilde{r}^{1}_{11}-\widetilde{r}^{04}_{1-1})\}}.
\label{r-ds-appr}
\end{eqnarray}  
Equation $(\ref{ds-r-beta-prec})$ is exact and can be used to calculate correction to  $\widetilde{R}_D$ due to the contribution of $\beta_D$.
The ratio $\widetilde{R}_D$ approximates $\widetilde{R}$ better than $\widetilde{R}_U$ since the numerator in formula
 $(\ref{def-beta-u})$ for the small parameter $\beta_D$ does 
not contain the contribution of 
$|U_{1\frac{1}{2}1\frac{1}{2}}|^2$, while the correction to $\widetilde{R}_U$ contains the contribution of 
$|U_{1\frac{1}{2}1\frac{1}{2}}|^2$. 

In the SCHC approximation, when all amplitudes with $\lambda_V \neq \lambda_{\gamma}$ are assumed to be negligibly small at small $v_T$  
one has to put 
$\widetilde{r}^{1}_{00}$, 
$\widetilde{r}^{1}_{11}$, $\widetilde{r}^{04}_{1-1}$ equal to zero in
Eq.~$(\ref{r-ds-appr})$ according to the formulas for $\widetilde{r}_{\lambda_V \tilde{\lambda}_V}^{\alpha}$ presented in \cite{SW,DC-24}. This leads to the relation 
$\widetilde{R} \approx \widetilde{R}_{04}$ less 
precise at high energies than $\widetilde{R} \approx \widetilde{R}_D$, where
\begin{eqnarray}
 \widetilde{R}_{04}
=\frac{\widetilde{r}^{04}_{00}}{\epsilon (1-\widetilde{r}^{04}_{00})}.
\label{r04-appr}
\end{eqnarray}
This statement will be proved below.
The quantity $\widetilde{R}_{04}$ is commonly used (see, for instance, \cite{Critt,INS,SW,H1-Aar,DC-24,DC-95,DC-95-err}) to estimate the ratio $\widetilde{R}$ from the experimental 
data. 
We note, that the improvement which gives formula $(\ref{r-ds-appr})$ in comparison to $(\ref{r04-appr})$ will be important for more precise data than the modern ones. Even at 
intermediate energies the difference $\widetilde{R}_{04}-\widetilde{R}_{D}$ is of the order of the total experimental uncertainty. Let us consider $\widetilde{R}_{04}$ for three 
bins
in $Q^2$ with mean values $<Q^2>=1.19$ GeV$^2$, $1.66$ GeV$^2$, and $3.06$ GeV$^2$ for the proton target from Ref.~\cite{DC-24}. The values of $<\epsilon>$ are $0.831$, $0.825$, 
and $0.790$, while the ratios $\widetilde{R}_{04}$ are $0.701 \pm 0.063$, $0.798 \pm 0.080$, and $1.053 \pm 0.074$. The values of $\widetilde{R}_{D}$ $0.617$, 
$0.795$, and $1.004$ are systematically smaller than the values of $\widetilde{R}_{04}$ by $0.084$, $0.003$, and $0.049$ but the differences are comparable to the total 
experimental uncertainties of $\widetilde{R}_{04}$ within less than 1.3 of the standard deviation.    

The difference between the exact ratio $\widetilde{R}$ and the approximate quantity $\widetilde{R}_{a}$ for 
$a=D$ or $a=04$ can be expressed with the exact relation
\begin{eqnarray}
\widetilde{R}-\widetilde{R}_{a}=-\frac{\beta_{a}}{\epsilon}\frac{(1+\epsilon \widetilde{R}_a)^2}{1+\beta_{a}(1+\epsilon \widetilde{R}_a)},
\label{dif-r-appr}
\end{eqnarray}
where $\beta_{a}$ and $\widetilde{R}_{a}$  for $a=D$ are respectively defined by Eqs.~$(\ref{def-beta-u})$ and $(\ref{r-ds-appr})$, while for $a=04$ $\beta_{04}$ is
\begin{eqnarray}  
\beta_{04}=\widetilde{\sum}\bigl\{-2\epsilon |T_{10}|^2-2\epsilon |U_{10}|^2 
+|T_{01}|^2 +|U_{01}|^2\bigr\}/\widetilde{\mathcal{N}} 
\label{def-beta-04}
\end{eqnarray}
and $\widetilde{R}_{04}$ is given by Eq.~$(\ref{r04-appr})$.
A comparison of Eq.~$(\ref{def-beta-u})$ with Eq.~$(\ref{def-beta-04})$ shows that $\beta_{D}$ contains only the UPE-amplitude contribution (without 
the contribution of $|U_{1\frac{1}{2}1\frac{1}{2}}|^2$) 
in the numerator of  Eq.~$(\ref{def-beta-u})$ which goes to zero at $W \to \infty$, while $\beta_{04}$  contains  the NPE-amplitude contribution ($|T_{01}|^2$ and $|T_{10}|^2$)
in addition to the UPE-amplitude contribution  
in the numerator of  Eq.~$(\ref{def-beta-04})$, therefore $\beta_{04}$ remains, generally speaking, nonzero when $W \to \infty$. This means that $|\beta_{D}| \ll |\beta_{04}|$ 
at high energies of the virtual photon.  
\section{Conclusion}  
\label{sect-6}
The exact formulas for the helicity-amplitude ratios of the vector-meson production by a polarized lepton beam on spinless targets in terms of the vector-meson 
spin-density-matrix elements are obtained.
It is shown also that data for an unpolarized beam are insufficient to find single-valued solutions for the amplitude ratios. 
The fit of the angular distribution normalized to unity in the amplitude method needs 8 free real parameters (4 real and 4 imaginary parts of the complex 
amplitude ratios), while the number of free parameters $r_{\lambda_V \tilde{\lambda}_V}^{\alpha}$ in the SDME method is 23. 
The SDMEs are not independent, hence 23-8 constraint relations between SDMEs exist. Tests of the constraint relation validity  permit to estimate the precision with which the 
Monte-Carlo codes describe the detector properties. 
It is reasonable to use both the SDME and the amplitude method since the differences between SDMEs extracted directly from data and calculated from the amplitude ratios
permit also to estimate the quality of detector property description with the Monte-Carlo codes.
The exact formulas $(\ref{r-less})$ and $(\ref{r-less-upe})$ for the virtual-photon longitudinal-to-transverse differential-cross-section ratio, 
$R$ in terms of the SDMEs are found for  spinless targets.  The new approximate formula $(\ref{r-ds-appr})$ more precise for high energies than the standard 
Eq.~$(\ref{r04-appr})$ for the ratio $\widetilde{R}$ in term of the SDME $\widetilde{r}^{04}_{00}$ applicable to scattering off the nucleon targets is proposed.
\section*{Acknowledgements}
I would like to thank Prof.~O.L.~Fedin, Prof.~V.T.~Kim, Dr.~M.B.~Zhalov, Dr.~E.N.~Komarov, Dr.~I.B.~Smirnov and all other participants of the High Energy Physics Division 
seminar in Petersburg Nuclear Physics Institute for constructive discussion. I am also much obliged to Prof. V.R.~Shaginyan for advises that improved the manuscript.
 
\end{document}